\documentclass[aps,prl,preprint,superscriptaddress]{revtex4}
\usepackage{graphicx}

\UseRawInputEncoding
\usepackage{color}

\begin{document}
\title{Resonances involving integer magnons and spin-1/2 excitations in a magnetism modulated two-dimensional electron gas}

\author{Lin Zhang}
\affiliation{Department of Mathematics and Physics, North China Electric Power University, Beijing 102206, China}
\affiliation{State Key Laboratory of Superlattices and Microstructures (SKLSM), Institute of Semiconductors, Chinese Academy of Science, P.O. Box 912, Beijing 100083, China}

\author{Yi Wang}
\affiliation{State Key Laboratory of Superlattices and Microstructures (SKLSM), Institute of Semiconductors, Chinese Academy of Science, P.O. Box 912, Beijing 100083, China}
\affiliation{College of Material Science and Opto-Electronic Technology, University of Chinese Academy of Sciences, Beijing 100190, China}

\author{Shu-Yu Zheng}
\affiliation{Beijing National Laboratory for Condensed Matter Physics, Institute of Physics, Chinese Academy of Sciences, Beijing 100190, China}

\author{Li Lu}
\affiliation{Beijing National Laboratory for Condensed Matter Physics, Institute of Physics, Chinese Academy of Sciences, Beijing 100190, China}

\author{Chi Zhang}
\altaffiliation{Electronic address: zhangchi@semi.ac.cn}
\affiliation{State Key Laboratory of Superlattices and Microstructures (SKLSM), Institute of Semiconductors, Chinese Academy of Science, P.O. Box 912, Beijing 100083, China}
\affiliation{College of Material Science and Opto-Electronic Technology, University of Chinese Academy of Sciences, Beijing 100190, China}
\affiliation{CAS Center for Excellence in Topological Quantum Computation, University of Chinese Academy of Sciences, Beijing 100190, China}



\begin{abstract}
We conduct an experimental study of high-mobility two-dimensional electron gas (2DEG) in GaAs/AlGaAs quantum wells modulated by strong magnetism at an in-plane magnetic field ($B$).
The modulated $B$-fields are performed via the single stripe and gratings which are made of the heavy rare earth metal Terbium (Tb) thin films on the sample surface.
Robust ferromagnetic resonances (FMRs) persist from the temperature of 1.5 K until up to 50-70 K in both the stripe and grating samples, for the ferromagnetism (FM) phase of Tb exists at above 100 K.
The high-order (with integer numbers of $j \equiv \hbar\omega/g\mu_{B}B = 1, 2,$...) magnetic resonances can also be observed in the stripe structure via the microwave (MW) photovoltaic detection and magnetoresistance under microwave irradiation.
It indicates that each absorbed photon in the resonance is accompanied by the excitation of an integer number of magnons.
In addition, the resonance features around $j = 1/2$ are robust in the single-stripe modulated sample, which suggests the spinons with spin-1/2 collective excitations in a 1D Heisenberg model.
Our experimental study provides a promising platform for discovering and manipulating quasiparticles and collective excitations (magnons and spinons) in magnetic phases or quantum devices.
\end{abstract}

\maketitle
\emph{Introduction.---}

The high-mobility two-dimensional electron gas (2DEG) in semiconductors provides a ideal platform to study ample quantum phenomena and electron states, including the quantum Hall states ~\cite{Klitzing1980, Tsui1982} and the non-equilibriums states under microwave manipulations ~\cite{Mani2002, Zudov2003}.
But due to the limit of the MBE growing of III-V and IV group semiconductor heterostructures, spin related transport and spectroscopic experiments involving collective excitations have not developed rapidly in the high-mobility 2DEG.
In the past few years, hybrid structures of magnetism modulated 2DEG have become an alternative plan.
In a magnetic field modulated 2DEG, various physics phenomena related to the spin in electron systems have been revealed, and the spatially modulated magnetic field plays an important role in electric driven spin resonance ~\cite{Nogaret2007}, magnetically modulated quantum Hall edge modes ~\cite{NogaretPRBrapid2017}, and helical magnetic modes dependent on Chern numbers ~\cite{Haldane1995, Young2014}.
Recent studies have discovered topological surface states in topological insulators ~\cite{Hasan-Kane} and nontrivial states in topological semimetals ~\cite{Xu2015}.

As a type of fundamental quasiparticles (or excitations), magnons involving spin waves in different magnetic resonance measurements, present the microscopic orders and dynamics in different magnetic phases via distinct quantum characteristics and collective behaviors.
And theoretical and experimental studies of magnons are prosperous in two-dimensional materials (e.g. single layer and twisted bilayer graphene under a perpendicular $B$) ~\cite{Wei2021, Alavirad2020, Zhou2022}, and ferromagnetic or antiferromagnetic matters ~\cite{Griffiths1946, Rodrigue1960, Klingler2015, Lee2016}.
The potential application of magnons based devices has been developed in both theories and models ~\cite{Meier-Loss, BinWei2021, Ke2023}.
Moreover, integrating magnons that is manipulated by microwave can be expected in the development of quantum information ~\cite{Jiang2023}.

Spinon is proposed as an important quasiparticle with charge neutral excitations, which occurs during the charge-spin separations, e.g. in the 1D Tomonaga-Luttinger liquid (TLL) model ~\cite{Jompol}.
Spinon is widely discussed in the theories of quantum spin liquid and strongly-correlated electron systems, but the experimental results are much less ~\cite{Bera2017}.
Because of the charge neutral characteristics, the spinon is very difficult to detect via electronic transport.
In our proposal, the spinons will be transformed into the charged particles through the hybrid junction via the spin pumping effect, and become electronic detectable in 2DEG ~\cite{Sandweg, Song2016}.

The FMRs under MW and transport are revealed in quantum wires modulated 2DEG ~\cite{NogaretJPCM2010}.
In the further study of 2DEG with the modulated prism magnets of small sizes (at about 100 nm scales) on the surface, magnetic dipolar interaction can be detected in photovoltage (PV) measurements ~\cite{NogaretJPCM2010, SaraivaPRB2010}.
In this paper, we report the transport of strongly magnetically-modulated 2DEG at a small in-plane $B$ with MW manipulations.
The single wire and grating structures that are made of the rare earth element Tb are deposited on the surface of embedded 2DEG.
The transport and MW photovoltaic measurements of our FM-modulated 2DEG are affected by spin wave transmission in proximity.
We demonstrate that 2DEG as a sensor can detect a series of magnetic resonances under an in-plane magnetic excitation, which involves an integer number of magnons in transport.
This finding accomplishes the manipulation of magnons in the electron system.
In addition, in the Tb single wire modulated sample, we observe strong resonance features at twice of the effective magnetic fields, which suggests the spinon excitations in 1D.
Due to the high Curie temperature of Tb, the observed magnetic resonances persist until around 50 to 70 K.

\emph{Method.---}
Our electron channel in the wafer has a high electron mobility $\mu =3.0 \times 10^{6} $cm$^{2}$/Vs and a density $n_{s} = 1.6 \times 10$$^{11}$ cm$^{-2}$ in the MBE grown modulation-doped GaAs/AlGaAs heterostructure, which is about 130 nm below the surface.
In our experimental proposal about the hybrid devices, the 2DEG channel is used as the detector of the collective spin excitations in the ferromagnetic matters via the microwave PV measurements by means of the spin pumping effect ~\cite{Sandweg, Song2016}.
The thin ferromagnetic metal structures are fabricated on (20 $\mu$m long and 10 $\mu$m wide) Hallbars.
The magnetic modulation generates a stray magnetic field, the perpendicular component of which deflects the ballistic trajectory in the 2DEG, coupling the electrical properties of the 2DEG with the magnetic properties of the grating ~\cite{NogaretJPCM2010}.
We conduct the measurements in two types of structures: one has a magnetic strip with a width of $d_{A} = 650$ nm on the surface of the Hallbar (sample A), and the other has nine magnetic gratings with a width of single wire $d_{B} = 500$ nm fabricated on the Hallbar with a pitch period of $a = 1000$ nm (sample B).
To avoid the edge effects, we fabricate the wire or the stripes with the length of $l = 60$ $\mu$m which is much longer than that of mesa in Hallbar (20 $\mu$m).
The single stripe structure on the surface of sample A is shown in Fig. 1, and the gratings on sample B are shown in Fig. 4.
The single stripe sample enables the magnetic modulation of electrons in a quasi-one-dimensional (quasi-1D) model.
The design of magnetic gratings is characterized by the dipole interaction between adjacent strips, and the aim is to investigate the effect of the interaction in the magnetic modulation process of 2DEG.
The mesas of the Hallbar are fabricated by UV lithography.
The patterns of the single wire and stripes are defined by the Raith-150 E-beam lithography (EBL), and the Tb thin film are evaporated via the ultrahigh vacuum (UHV) magnetron sputtering (for ferromagnetic metals) with a base vacuum of 10$^{-8}$ Torr.

The magnetism of the rare earth elements (or rare earth metals) stems from the RKKY-like exchange between the 4f conduction electrons and the local spins ~\cite{Cooke1979}.
The magnetic moment of the Tb atom is very large ($m_{Tb} \sim 9.0$) ~\cite{Lide}.
The electron group state of Tb is labeled as ($^{6} H_{15/2}$) with a spin angular momentum quantum number (q.n.) $S = 5/2$, the orbital angular momentum q.n.of $L = 5$, and the total angular momentum q.n. $J = 15/2$.
In our experiments, the electron channel is coupled to the Tb thin film on the surface while modulated by the spin wave stimulated by the MW ~\cite{NogaretJPCM2010}, and the resonances related to spin modes can be detected in the PV and transport study.

Our experiment is performed in a VTI with a base temperature of 1.5 K and a vector magnet of 8 T/ 2 T.
The PV measurements are carried out by chopping the MW radiation at 870 Hz with a lock-in amplifier without excitation current ~\cite{Mi2016, JFZhang2020}.
The high-resolution MW photovoltaic effect presents a series of discrete magnetic resonances at small $B$-fields.
An external in-plane $B$ field ($B_x$ or $B_y$) is applied to magnetize the Tb stripe (as shown in the inset of Fig. 1(a)).
The PV is detected via the modulated MW source by means of lock-in amplifiers.

\emph{Experimental Results.---}

Ferromagnetic resonance is an efficient tool to probe spin waves which involves magnons.
The motion of magnetization can be simplified by the Landau-Lifshitz-Gilbert equation:
\begin{equation}\label{1}
    \frac{\partial \overrightarrow{M}}{\partial t}= - \gamma (\overrightarrow{M} \times \overrightarrow{H_{eff}}) + \frac{G}{\gamma M_{s}^{2}} [\overrightarrow{M} \times \frac{\partial \overrightarrow{M}}{\partial t}]
\end{equation}
which includes two sections: the first term represents the precession, and the second term expresses the viscous damping (with a Gilbert constant $G$) that is used in the description of relaxations.
In various measurements, the resonance resembles a Lorentzian lineshape, and the linewidth is related to the relaxation process.
The resonant field is dependent on various factors, e.g. the anisotropy, $g$-factor, and magnetization.

In experiments, the FMR features can be detected via the magnetic susceptibility measurements in resonators or the photovoltaic probes.
The power is absorbed by the processing magnetization of the material.
In an FM phase, the frequency of the spin precession induced by the magnetization is given by the relation in ~\cite{Kittel1948}:
\begin{equation}\label{2}
    \omega = \gamma \mu_0 \sqrt{[H_{dc}^{e} + (N_{x}-N_{y})M_{dc}] [H_{dc}^{e} + (N_{z}-N_{y})M_{dc}]}
\end{equation}
, where $\gamma = g \mu_B / \hbar$ is the gyromagnetic ratio, $M_{dc}$ represents the magnetization induced by a static $B$-field, and $H_{dc}^{e}$ represents an effective magnetic field caused by the dc component of the applied in-plane $B$ ~\cite{SaraivaPRB2010}.
The demagnetization factors $N_{x}$, $N_{y}$, and $N_{z}$ are solely dependent on the geometry of the sample ~\cite{Kittel2004, Gurevich1996}.
Suppose the relation $ M =\chi H$, Eq. (2) can be simplified as:
\begin{equation}\label{3}
     \omega = \gamma \mu_{0} H_{dc}^{e}\sqrt{(1 - N_{y}\chi) (1 - N_{y}\chi + N_{z}\chi)}
\end{equation}.
In our samples with FM metallic rectangular prisms, the geometric parameters $N_{x} = 0$, and $N_{y}$ and $N_{z}$ can be calculated for prototype geometries ~\cite{Chen2005}.
FMR occurs in our samples and the MW frequency is proportional to the external $B$-field.
In a FM phase the relation $B = \mu_{0} H$ is effective at a low $B$-field.
So Eq.(3) can be further simplified as $\hbar \omega = \hbar \gamma (\mu_{0} H) A_{g} \approx g \mu_{B}B$, where the geometry factor is defined as: $A_{g} \equiv \sqrt{(1 - N_{y}\chi) (1 - N_{y}\chi + N_{z}\chi)}$, which is very close to 1 for the rectangular prism geometry.

Moreover, the slope of sample A ($\Delta \mu_{0} H_{dc}^{e})/\Delta f$ is about $\sim 0.519$ kG/GHz, and that of sample B is about $\sim 0.522$ kG/GHz.
Based on Eq. (3), the magneto-crystalline anisotropy results in an internal magnetic field, leading to an $\overline{H_{h}}$ included in $H_{a}$ ~\cite{Kittel1948}.
In our analysis, the crystal-field anisotropy of Tb exhibits a shift field of $H_{a}^e=H_{a}-\overline{H_h}$.

We conduct PV measurements on 2DES with the Tb wire (or gratings) structure on the surface.
As the Tb wire (or gratings) is magnetized along the $y$-axis, magnetic poles form on the edges of the stripe(s) along the width, generating a spatially varying $B$-field with two components: along the $z$-axis, and along $y$ (or $x$ orientation) which is parallel to the applied $B$-fields.
The $z$-component of the $B$-field that induces eddy currents transfers magnetization to the 2DEG via the high-frequency electromagnetic wave, thus forms a series of dips in the PV.
In our PV detection (PV vs. ($B-\mu_{0}H_{C}$)) in sample A, the FMR features at $T = 1.5$ K are robust at high power ($P \sim 20$ dBm) microwave and at frequencies from 1 to 18 GHz which is linear to the occurring in-plane $B$ along the $y$-axis.
Figure 1(a) displays the experimental data of microwave PV (at $f = 10$ GHz) and which display a series of resonance features superposed with a background increasing with the $B$-field.
Our PV curve exhibits a parabolic envelope, which is consistent with the PV curves at various temperatures below 70 K (as shown in Fig. 3(a)).
The resonance features decay much faster with temperature than the envelopes do, so the contribution from the 2DEG and the FM metals can be separated in our PV signals.
Thus we subtract the parabolic profile from the PV curve and obtain the trace $\Delta$PV vs. in-plane magnetic fields.
As shown in the PV trace extracted the background ($\Delta$PV) in Fig. 1(b), FMR occurs at ($B_{y}-\mu_{0}H_{C}$) around 5.1 kG at $f = 10$ GHz MW.
The resonance absorptions in PV detection are robust near $B_{y}-\mu_{0}H_{C} = 5.1$ kG and 10.2 kG.
To our surprise, the amplitude at $B_{y}-\mu_{0}H_{C} = 10.2$ kG that corresponds to $j = 1/2$ is much stronger than that of the FMR $j = 1$.
Because the FMR is accompanied by the magnons whose (bosonic) excitation exists in both 2D and 3D.
In the theory of 1D ferromagnetic chain of Heisenberg model ~\cite{Meier-Loss}, the term in Hamiltonian $g \mu_{B} B <s_{z}>$ provides a plausible explanation: the spinon transition occurs along with the spin-1/2 collective excitations (spinons) which are neither the fermions nor the bosons in 1D confinement ~\cite{Bera2017}.

In addition, a series of magnetic resonances in PV, i.e., near $B_{y}-\mu_{0}H_{C} = 2.55$ and 1.61 kG, are observable, which correspond to the resonances at $j = 2$ and 3 respectively.
Moreover, minor resonances are detected at very low $B$-fields.
The small dips at low $B$ (below 2 kG) suggest that the spin wave modes oscillate within the magnetic stripe well ~\cite{SaraivaPRB2010}.
The spin wave modes that propagate transversely inside the stripe become confined within the region of the lower internal field ~\cite{Bayer2006}.
In our sample with a large width ($w$) of a single Tb wire, the local spin wave mode (i.e., the dipolar electron spin wave (DESW)) that stems from the geometry confinement ~\cite{SaraivaPRB2010} is weakened to some degree in the PV experiments.
So the magnetic resonances that involve the integer magnons dominate the dipolar spin wave modes at very low in-plane $B$.

In order to confirm the phenomenon, we present the zoom-in features of the resonances involving integer magnons, as shown in Fig. 1(c).
The minima of $j=1, 2, 3...$ can be observed in both the $B$ sweeping up (blue color) and sweeping down (red color) traces.
The jumps in PV at around $j = 1, 2$ occur at high power MW (with $P= 20$ dBm), which are attributed to the effect from high power MW ~\cite{Gui2009}.
In addition, we hightlight the positions at $j = 4, 5$ with arrows, most minimal features are observable in both curves and only a minima at $j=4$ smears out in the (red color) sweeping down curve.

In our MW $f$-dependent PV detection with a frequency range of $\sim 4 - 16$ GHz, a series of magnetic resonances occur at integer $j$, which is defined as $j \equiv \hbar \omega /(g \mu_{B} B A_{g}) \approx \hbar \omega /(g \mu_{B} B)$, where $A_{g}$ is the geometry factor expressed in Eq.(3) and $A_{g} \sim 1$.
The PV signals (of sample A) at $f = 4, 10, 14$ GHz and at a sweeping down $B_{y}$ are shown in Fig. 2(a), 2(b), and 2(c), respectively.
At 4 and 10 GHz resonances occur at $j= 1$ and 1/2, which are marked by the black and red arrows.
But in the PV curve at 14 GHz in Fig. 2(c), the solid black arrows and dashed blue arrows highlight the resonances at $j = 1$ and 2 respectively, for the location of $j = 1/2$ is beyond the limit of in-plane $B$ of the magnet in the VTI.
A prototype FMR (at $j =1$) occurs at a higher magnetic field which is proportional (or linear) to the excited MW frequency (e.g., $B_{y}-\mu_{0}H_{C} = 2.13$ kG at 4 GHz, and $B_{y}-\mu_{0}H_{C} = 7.31$ kG at 14 GHz).
The data points of $B_{y}-\mu_{0}H_{C}$ vs. frequency illustrate a distinct slope of $\mu_{0} \Delta H_{dc}^{e}/(\Delta f) = 0.507$ kG/GHz for sample A, and $\mu_{0} \bar{H_{h}} = 0.26$ kG, as shown in Fig. 2(d).
And the effective $g$-factor can be obtained as approximately $g \sim 1.9$, which is very similar to those of Co and Dy ~\cite{SaraivaPRB2010}.
The magnon transmission is perpendicular to the applied $B$, e.g., the transmission of magnons along the $x$-direction with an external $B$ along the $y$-axis.

The $f$-dependent PV results at a $B$ along the $x$-direction of sample A are shown in Fig. 2(e) and 2(f), the FMRs (at $j = 1$) are still observable ($B_{x}-\mu_{0}H_{C}$ is around 4.7 kG at $f = 9$ GHz, and is around 7 kG at 14 GHz).
It is distinct from the experimental results of Dy modulated 2DEG ~\cite{NogaretEPL2011}, which is explained by the invalid magnetic modulation in the configuration: due to the shape of the slim rectangular prism with $l \gg d$, the magnon transmission along the length ($l$) leads to much weaker scattering than that along the width ($d$) direction ~\cite{Meier-Loss}.
But our observation of FMR at $B$ along the $x$-axis does not support the analysis.
One plausible reason is that the diffusion length (or the correlation length) of the magnons is longer than that of adjacent Hallbar arms and the width of Hallbar.

Among all heavy rare earth elements, Tb has one of the highest Curie temperature $T_{C} \sim 220$ K ~\cite{Lide}, which is valuable in the potential application of the hybrid devices.
At the base-$T$ around 1.5 K, the magnetic resonances remain robust with the Tb stripe magnetized along the $y$-direction, as shown in Fig. 3(a).
At the high $T$ range of 50 -70 K, the magnetic resonances (at various integer $j$) are robust at 50-70 K at ($B_{y} - \mu_{0} H_{C}$) $\sim 5.2$ and 10.0 kG.
Moreover, the prime magnetic resonance (FMR at $j = 1$) persists until around 100 K with the magnetization along the $y$-axis, and finally disappears at 150 K.

For comparison, we detect the $T$-dependent resonances at a $B$ along the $x$-direction, as shown in Fig. 3(b).
The FMRs at $f= 9$ GHz remain robust until up to 30 K, and the features become very weak at $T \sim 50$ K.
And the transitional temperatures of the FM disappearance in 2DEG are much lower than those at the transitional point of Terbium.
But the high-order magnetic resonances $j = 2, 3$ cannot be observed in our $T$-dependent PV detection at $B_{x}$.

Magnetic resonances in 2DEG with an FM grating structure exhibit more complex magnetic excitations than the single stripe structure, which results from the dipolar magnetization waves (DMW) propagating between the nearest neighboring stripes ~\cite{SaraivaPRB2010}.
But at $f = 9, 11, 15$ GHz MW (in Fig. 4(a), 4(b), and 4(c), respectively), no features of DMW are detectable in our device (sample B).
A plausible reason is that the distance between the nearest neighboring Tb wires is not small enough, so that the coupling between the neighboring wires is relatively weak and the DMW mechanism is dominated by the magnetic resonance features.

Figure 4(d) shows the PV and resistance ($R_{xx}$) measurement results (of sample B) at a $B$-field along the $y$-direction under $f = 16$ GHz MW irradiation.
The magnetic resonance features at $j = 1, 2$ can be observed at the shoulders in PV and at around the peaks in resistance, which are highlighted by the black color arrows.

\begin{figure}
 \includegraphics[width=0.8\linewidth]{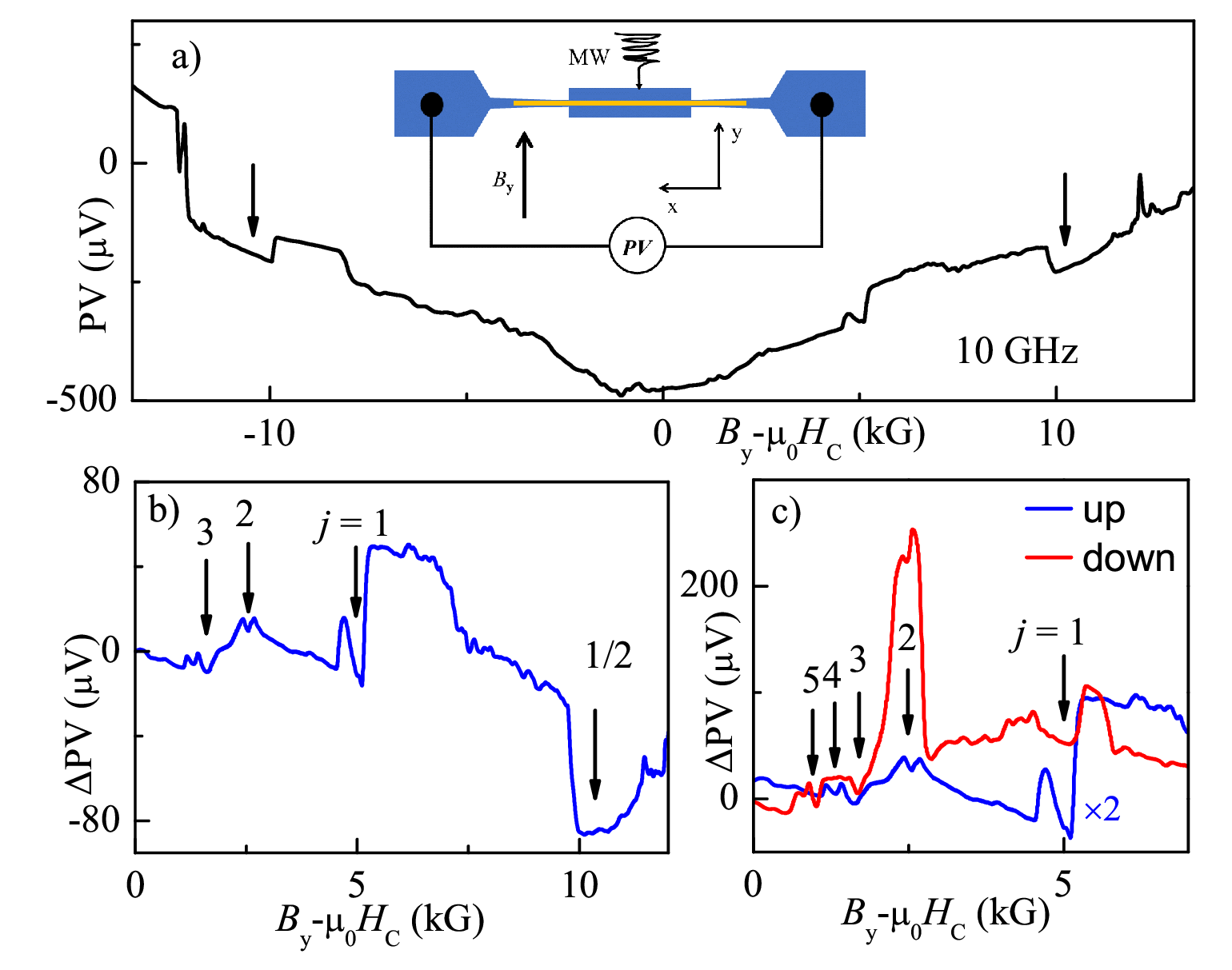}
 \caption{(Color online) Panel (a): Photovoltage (at 1.5 K) of sample A under 10 GHz/ 20 dBm MW irradiation (black curve).
 The inset display the MW PV detection of Tb-stripe-2DEG via a lock-in amplifier.
 The yellow bar with a width of 650 nm shows the 70 nm thick Tb film on the Hallbar.
 (b): For clarify the resonance features we removed the background of curve and obtained the $\Delta$PV trace, where the robust resonance at $j \sim 1, 2$ and 1/2 are marked by the arrows.
 (c): In the comparison of sweep-up and sweep-down data, the resonance features involving integer magnons are detectable.
 }
  \label{FIG1}
 \end{figure}

\begin{figure}
 \includegraphics[width=0.8\linewidth]{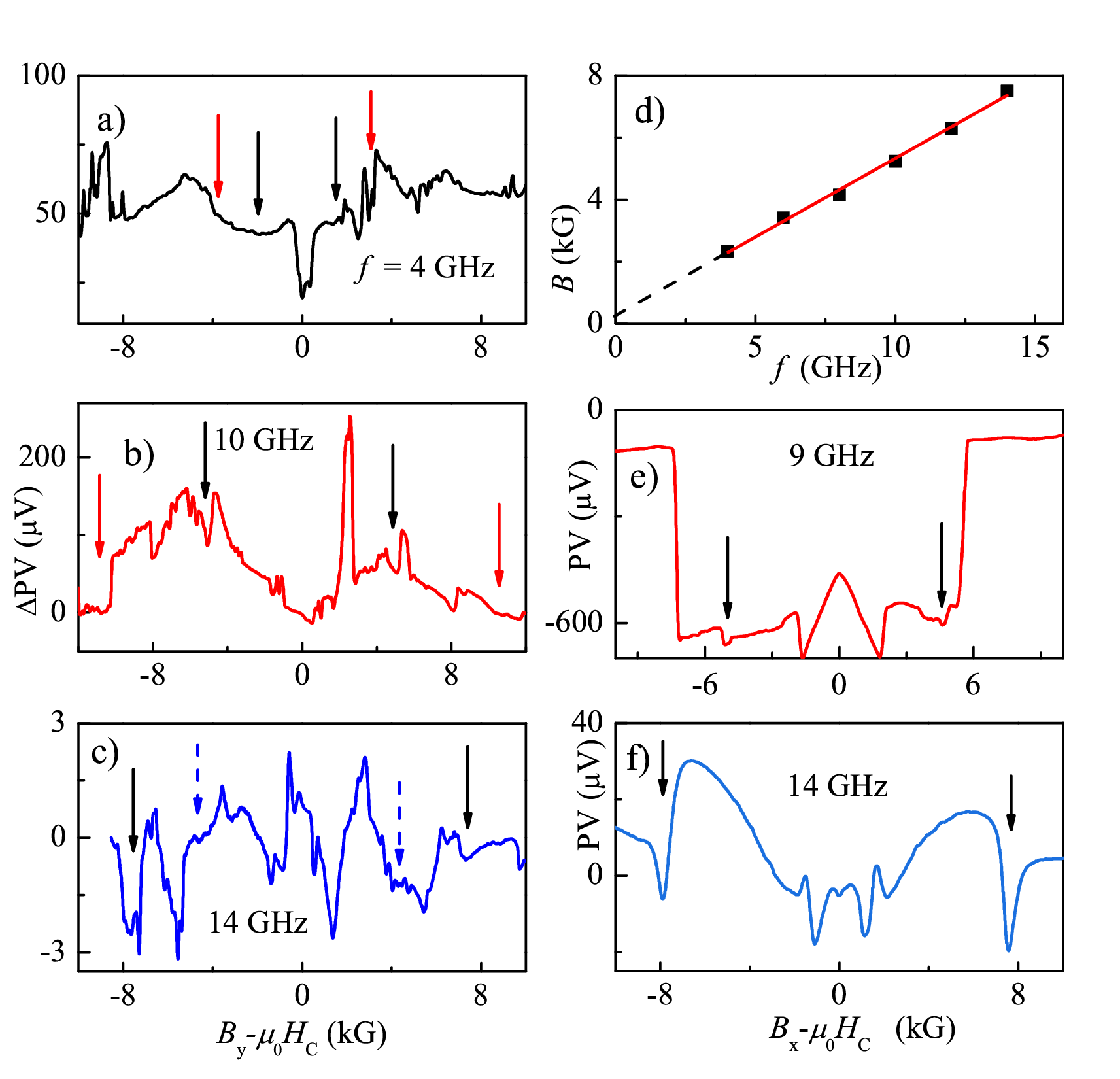}
 \caption{(Color online) Panel (a), (b), (c): FMRs (at 1.5 K) of sample A at $f = 4, 10, 14$ GHz and 20 dBm MW and at an in-plane $B$ along y-axis ($B \parallel y$), which are highlighted by the arrows.
 (d): MW frequency ($f$) versus ($B_y-\mu_{0}H_{C}$) in FMR ($j = 1$).
 (e), (f): FMRs at 9 and 14 GHz and at $B \parallel x$ (PV vs. ($B_{x} - \mu_{0}H_{C}$)), the $j = 1$ features are highlighted by the arrows.
 }
  \label{FIG2}
 \end{figure}

\begin{figure}
 \includegraphics[width=0.8\linewidth]{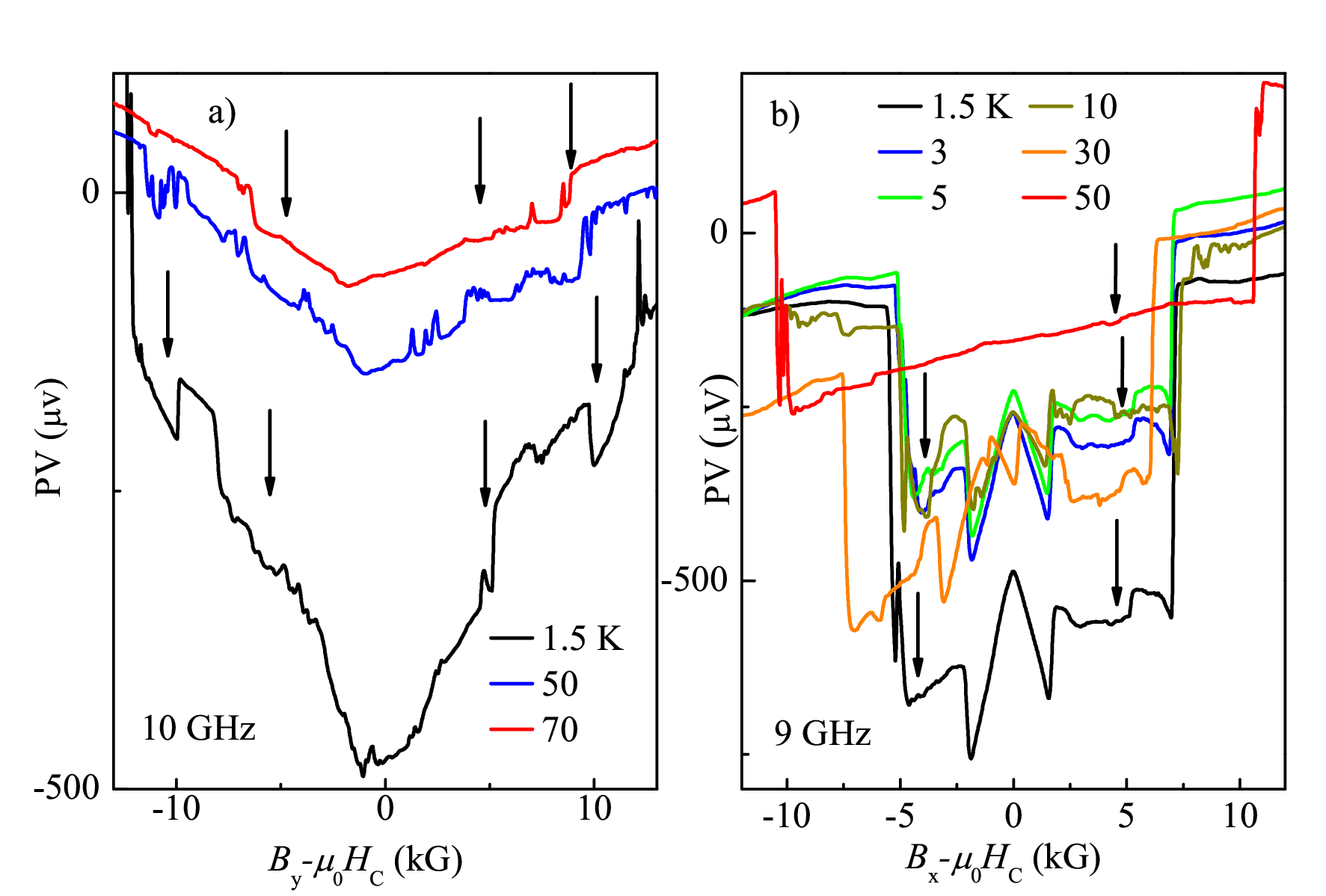}
 \caption{(Color online) Panel (a): $T$-dependent PV of sample A at f = 10 GHz and 20 dBm MW with applied $B$ long $y$-axis ($B \parallel y$).
 The FMR persistent at 70 K.
 (b) $T$-dependent PV of sample A at 9 GHz and 20 dBm MW and $B$ along $x$-direction ($B \parallel x$).
 }
  \label{FIG3}
 \end{figure}

\begin{figure}
 \includegraphics[width=0.8\linewidth]{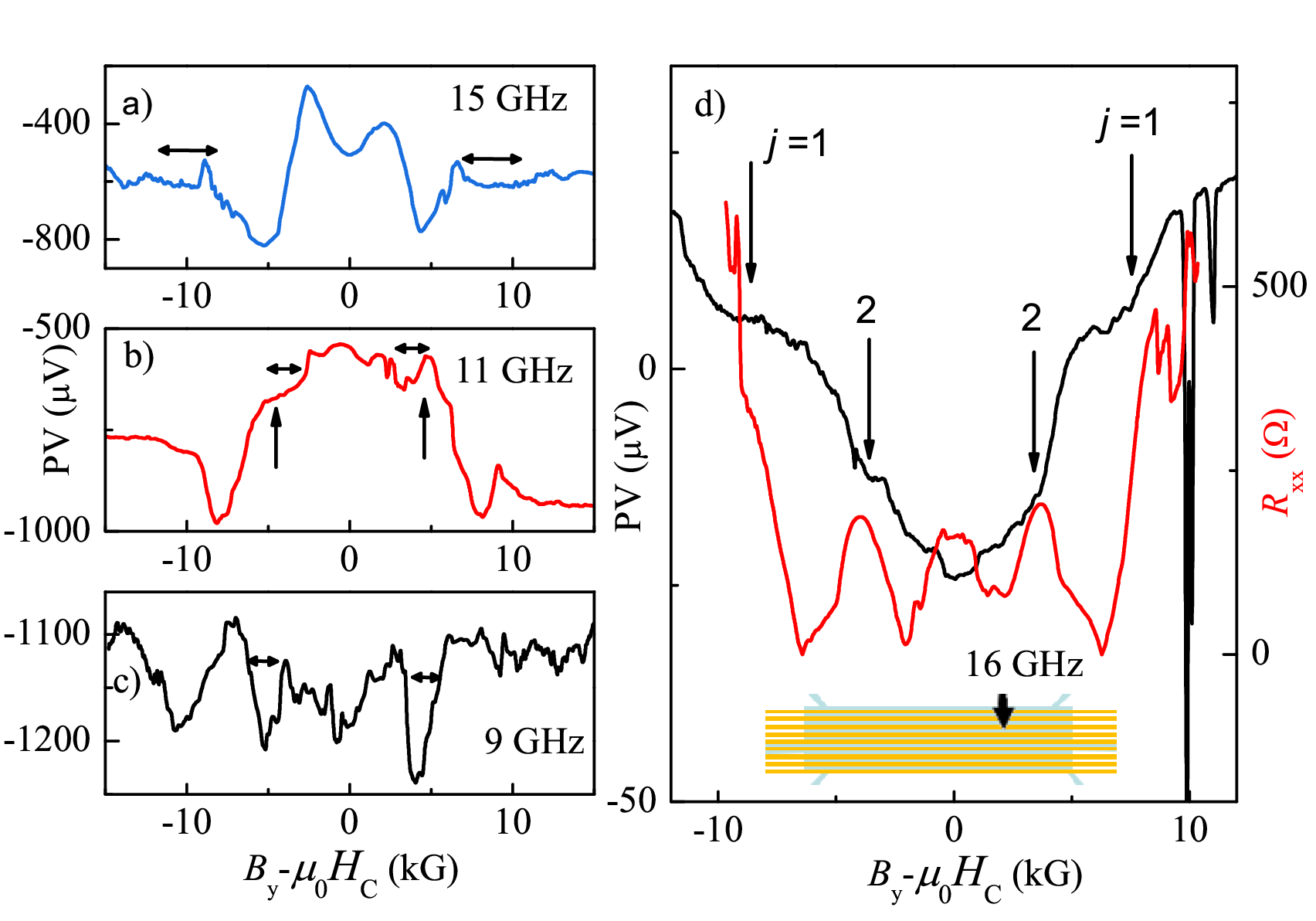}
 \caption{(Color online) Panel (a), (b), (c): FMRs (at 1.5 K) of the (gratings) sample B in the PV detection at $B \parallel y$), with the frequencies of 15, 11, 9 GHz, respectively.
 (d): The $f = 16$ GHz PV and the resistance under MW irradiation (at $f =16$ GHz), at a $B \parallel y$.
 Inset: the schematic diagram of the grating structures on sample B.
 The yellow bars on the Hallbar displays the 70 nm thick Tb gratings.
 }
  \label{FIG4}
 \end{figure}

\emph{Discussion.---}

(1) We observe robust FMRs in both single wire and grating patterned samples.
The linewidth of the resonances is directly relevant to the relaxation processions.
In our observation, the FWHM of FMR in the 1D single FM wire sample does not exhibit obvious $B$-dependence at low $T$.
But the FWHM of FMR increases with the occurring $B$-field (as shown in Fig.4).
Because the width of FMR increases with the dipolar magnetic fields which involves the dipolar mode between the adjacent rectangular prisms (i.e. the DMW) ~\cite{SaraivaPRB2010}.

(2) In the 2DEG sample patterned with single wire or grating structures, the local spin wave mode includes DESW and DMW which originate from the dipolar interaction (or effect) within the single wire or between the nearest neighboring wires ~\cite{SaraivaPRB2010}.
In our sample with a large width of wire or gap between the adjacent stripes, DESW and DMW are weakened to a great extent.
And the magnetic resonances that involve integer magnons dominate the local spin wave modes.

(3) The comparison of our measurements between single wire and gratings structures shows differences in the excitations around $j = 1/2$.
The robust excitations of $S=1/2$ can be detected only in the 1D single wire-modulated 2DEG, but cannot be observed in the 2DEG sample with gratings structures.
So in the 2DES with 1D FM-modulation, the occurrence of excitations at $j=1/2$ suggests the existence of spinons with $S = 1/2$.
As we know, the magnetic resonance transitions involve magnons, for instance, the spin wave of the FMR in the microwave PV measurements in the 2D or 3D models require the integer transition of $<S_{z}>$, because magnon is a typical boson.
So in the 2DEG with gratings structure, the 1D characteristics of the FM-modulation is substantially weakened.
However, in the real 1D models, the rule of the boson or fermion absorption is invalid, and the excitation of spinons is a very plausible explanation.

(4) In our PV and transport measurements, the high-mobility 2DEG plays a role as the sensor for detecting of the magnons and spinons.
Since the quasi-particles or excitations of spinons are charge neutral, which cannot be detected by means of PV or transport.
But due to the spin pumping effect in the the FM metal-2DEG or topological insulator junctions ~\cite{Song2016}, the charge neutral spin signals turn into the charged signals and the measurements of the collective excitations with spin-1/2 become feasible.

(5) Ideally, the Tb thin film appears to be a ferromagnetic phase below $T = 220$ K.
And we observe distinct resonance features at $j = 1/2$, which correspond to the transition with a collective excitation of $S = 1/2$.
In comparison, a recent experimental study revealed the excitation with spin-1/2 in the antiferromagnetic (AFM) chain of 1D Heisenberg model ~\cite{Bera2017}, which differs from our study in the ferromagnetic chain.
It seems likely that the spin-1/2 excitations occur in AFM or ferrimagnetic chain in 1D models.
So the occurrence of the polarized FM chain at a $B$-field with spin-1/2 excitation is reasonable, although the details of the mechanism remain to be explored.

In addition, the resonances at $j = 1/2$ can be explained in the frame of spin-1/2 in spinon liquids ~\cite{Yu2022}.
However, in our $T$-dependent study the resonance features at $j = 1/2$ persist until around 50 K (about 4.6 meV) which is much higher than the Fermi energy in the channel.
Thus the collective behavior (such as the FMR in the electron channel) is the only reasonable explanation, and the ESR mechanism can be ruled out in our study.

\emph{Conclusion.---}


In summary, we have conducted a systematic study of quasi-1D ferromagnetism modulated 2DEG devices via the electrical and microwave PV detections, and first revealed a series of magnetic resonances associated with integer magnons (with $j = 1, 2, 3,...$).
The phenomenon is not only clearly demonstrated in the Tb-single wire structure hybrid device but also effectively verified in the grating structure sample.
Surprisingly, in our quasi-1D hybrid device, the most robust resonance occurs at $j = 1/2$ which is more intensive than FMR (at $j = 1$).
It suggests that the spinon excitations with spin-1/2 are effective in the 1D Heisenberg model.
Due to the spin pumping effect, the charge neutral spinons transform into charged particles that are electronically measurable in the 2DEG channel.
Moreover, all the quasiparticles or collective excitations (including integer magnons and spinons) persist until up to 50 K.
So the transmission study and precise manipulation of a finite number of magnons can be utilized to develop magnon-based quantum devices.
And integrating magnon in quantum information can be expected to be fulfilled at relatively high temperatures that are very close to the liquid Nitrogen temperature.
The observation of spin-1/2 resonances that is associated with collective excitations of spinons in the 1D ferromagnetic Heisenberg model, will stimulate further discovery in both theory and experiment in various quantum systems.

\begin{acknowledgments}
This project is supported by the National Science Foundation of China (Grant No.11974339), and by the Strategic Priority Research Program of the Chinese Academy of Science (Grant No. XDB 0460000).
L.Z., Y.W. and C.Z. performed the experiments; L.Z. and C.Z. analyzed the data; L.Z drafted and C.Z. wrote the paper; L.Z. and Y.W. carried out the cleanroom work; S.Z. and L.L. grew the high-quality semiconductor wafers; C.Z. conceived and supervised the project.
\end{acknowledgments}

\end{document}